\documentclass[12pt]{article}
\usepackage{graphicx}
\usepackage{latexsym}
\def\be{\begin{equation}}
\def\ee{\end{equation}}
\def\bea{\begin{eqnarray}}
\def\eea{\end{eqnarray}}
\begin{document}

\begin{flushright}
5th March 2007\\
hep-th/0703040
\end{flushright}

\begin{center}
{\Large \bf Ekpyrotic collapse with multiple fields}
\vskip 1cm

Kazuya Koyama \footnote{ E-mail: Kazuya.Koyama@port.ac.uk}
 and
David Wands \footnote{E-mail: David.Wands@port.ac.uk }

\vskip 1cm Institute of Cosmology and Gravitation, Mercantile House,
University of Portsmouth, Portsmouth~PO1~2EG, United Kingdom
\end{center}

\begin{abstract}

A scale invariant spectrum of isocurvature perturbations is
generated during collapse in the scaling solution in models where
two or more fields have steep negative exponential potentials. The
scale invariance of the spectrum is realised by a tachyonic
instability in the isocurvature field. We show that this instability
is due to the fact that the scaling solution is a saddle point in
the phase space. The late time attractor is identified with a single
field dominated ekpyrotic collapse in which a steep blue spectrum
for isocurvature perturbations is found. Although quantum
fluctuations do not necessarily to disrupt the classical solution,
an additional preceding stage is required to establish classical
homogeneity.

\end{abstract}

\section{Introduction}
Given the success of simple inflation models in explaining the
primordial homogeneity and almost scale-invariant spectrum of
curvature perturbations on large scales, it is important to consider
whether there is any other model for the early universe that might
provide an alternative explanation.
One contender is the idea of an early collapse phase rather than an
accelerated expansion. This invokes the same basic mechanism as
inflation to generate large-scale perturbations. A shrinking comoving
Hubble scale can take quantum vacuum fluctuations of scalar fields on
small scales and evolve them into overdamped perturbations on
super-Hubble scales. But the approach to solving the homogeneity
problem is quite different.

In pre big bang model of Gasperini and Veneziano
\cite{Gasperini:1992em} the collapsing universe (in the Einstein
frame \cite{Gasperini:1993hu}) is dominated by the kinetic energy of
massless fields in the low energy string effective action. Such a
solution is an attractor in a collapsing universe for a wide range
of interaction potentials so offers a solution to the homogeneity
problem, though this kinetic-dominated collapse is only marginally
stable with respect to anisotropic shear \cite{Kunze:1999xp}. By
contrast the ekpyrotic scenario proposed an ultra-stiff collapse
phase dominated by a steep negative potential \cite{Khoury:2001wf}
(see also \cite{Kallosh:2001ai,Khoury:2001iy}). This is a late time
attractor and is stable with respect to anisotropy.

On the other hand both models produce a steep blue spectrum of
perturbations (in the spatially flat gauge) in the fields driving the
collapse and hence in the comoving curvature perturbation
\cite{Lyth:2001pf,Brandenberger:2001bs,Hwang:2001ga,Khoury:2001zk}. A
collapse phase dominated by a field with a pressureless equation of
state does yield a scale-invariant spectrum of comoving curvature
perturbations \cite{Wands:1998yp,Finelli:2001sr,Allen:2004vz} but it is
not a late time attractor \cite{Heard:2002dr,Gratton:2003pe}.

It was shown in \cite{Copeland:1997ug} that axion fields in the pre
big bang can acquire a scale-invariant spectrum of isocurvature field
perturbations in a pre big bang phase and that this could in principle
be converted to curvature perturbations at a bounce
\cite{Lidsey:1999mc} or some time after the bounce via the curvaton
mechanism \cite{Enqvist:2001zp,Lyth:2001nq}. The idea of producing a
scale-invariant spectrum of isocurvature perturbations in axion fields
was also investigated in the ekpyrotic scenario
\cite{Notari:2002yc,DiMarco:2002eb}.

In the ekpyrotic scenario it is also possible to generate a
scale-invariant spectrum of isocurvature perturbations in models
where two or more fields have steep exponential potentials. In this
case there exists a scaling solution
\cite{Finelli:2002we,Guo:2003rs} where the energy densities of the
fields grow at the same rate during collapse, analogous to assisted
inflation solutions in an expanding universe \cite{Liddle:1998jc}.
Several authors
\cite{Lehners:2007ac,Buchbinder:2007ad,Creminelli:2007aq} have
recently proposed new ekpyrotic models based on this scaling
solution.

If a scale-invariant spectrum of comoving curvature perturbations
can be generated during the collapse phase then there is still the
``graceful exit problem'' of turning collapse to expansion. In the
pre big bang scenario it was envisaged that such a bounce could be
achieved by higher-order corrections to string effective action
which violate the null energy condition
\cite{Gasperini:1996fu,Brustein:1997ny,Brustein:1997cv}. A bounce
due to higher-order kinetic terms in a ghost condensate model has
recently been considered in the new ekpyrotic scenarios
\cite{Buchbinder:2007ad,Creminelli:2007aq} (see also
\cite{Creminelli:2006xe,Tsujikawa:2002qc}).

In this paper we discuss the tachyonic instability of long
wavelength perturbations about the new ekpyrotic scaling solution
which leads to the scale-invariant spectrum
\cite{Finelli:2002we,Lehners:2007ac,Buchbinder:2007ad,Creminelli:2007aq}. 
We show that the solution is
not an attractor at early or late times, but rather a saddle point
in the phase-space. The late time attractor is the old ekpyrotic
collapse dominated by a single field and the instability drives the
scaling solution to this late time attractor. By contrast we find
steep blue spectra for isocurvature field perturbations about the
kinetic dominated early time attractors, or single-field dominated
ekpyrotic late time attractors.
% DW
We present the phase space analysis in an Appendix.

\section{Homogeneous field dynamics}

During the ekpyrotic collapse the expansion of the universe is assumed
to be described by a 4D Friedmann equation in the Einstein frame with
scalar fields with negative exponential potentials
\begin{equation}
 3 H^2 = V + \frac12 \dot\phi_i^2 \,,
\end{equation}
where
\be
 \label{Vi}
 V = - \sum_i V_i e^{-c_i\phi_i} \,,
\ee
and we take $V_i>0$
and set $8\pi G$ equal to unity.

The authors of \cite{Lehners:2007ac} found a scaling solution (previously
studied in \cite{Finelli:2002we,Guo:2003rs}) in which both fields roll
down their potential as the universe approaches a big crunch
singularity, analogous to the assisted inflation dynamics found by
\cite{Liddle:1998jc} for positive potentials in an expanding universe.
In this assisted ekpyrotic collapse we find power-law solution for the
scale factor
\be
 \label{scalingp}
 a \propto (-t)^p \,, \quad {\rm where}\ p = \sum_i \frac{2}{c_i^2} \,,
\ee
where
\be
 \label{backgroundij}
 \frac{\dot\phi_i^2}{\dot\phi_j^2}
 = \frac{-V_i e^{-c_i\phi_i}}{-V_j e^{-c_j\phi_j}}
 = \frac{c_j^2}{c_i^2} \,.
\ee

As in the case of assisted inflation, this behaviour is easily
understood after a rotation in field space
\cite{Finelli:2002we, Malik:1998gy} into so-called adiabatic field,
$\sigma$, along the background trajectory (\ref{backgroundij}), and
the orthogonal, entropy field directions, $\chi_i$, \cite{Gordon:2000hv}.
The potential (\ref{Vi}) can then be simply re-written as
\be
 \label{Vsigmachi}
 V = - U(\chi_i) \, e^{-c\sigma} \,,
\ee
where
\be
\frac{1}{c^2} = \sum_i \frac{1}{c_i^2} \,.
\ee

To simplify the analysis we restrict ourselves to the case of two
scalar fields, though the general discussion applies to an arbitrary
number of fields. In this case we have
 \be
\sigma = \frac{c_2 \phi_1 + c_1 \phi_2}{\sqrt{c_1^2+c_2^2}} \,,
\quad
\chi = \frac{c_1 \phi_1 - c_2 \phi_2}{\sqrt{c_1^2+c_2^2}} \,,
 \ee
and we can write the explicit dependence of the potential
(\ref{Vsigmachi}) on the orthogonal field \cite{Malik:1998gy, Finelli:2002we}
\be
 U(\chi) = V_1\, e^{-(c_1/c_2)c\chi} + V_2\, e^{(c_2/c_1)c\chi} \,.
\ee
This is clearly a positive function (given $V_1>0$ and $V_2>0$)
and bounded from below. Close to its minimum we can expand this as
\cite{Malik:1998gy,Finelli:2002we}
\be
 U(\chi) = U_0 \left[ 1 + \frac{c^2}{2}(\chi-\chi_0)^2 + \dots \right] \,.
\ee

Thus we can confirm that there is a classical trajectory for the two
fields in which $\chi$ remains fixed, $\chi=\chi_0$, while the
adiabatic field $\sigma$ rolls down a steep exponential potential
(\ref{Vsigmachi}), which reduces to
 \be
 V|_{\chi=\chi_0} = - U_0 \, e^{-c\sigma} \,,
 \ee
and we identify the scaling solution (\ref{scalingp}) with the usual
power law solution for a single field solution for a negative
exponential potential which exists for $c^2>6$
\cite{Heard:2002dr,Gratton:2003pe}.

However we also see that the entropy field, $\chi$, has a negative
mass-squared
 \be
 m_\chi^2 = c^2 V < 0 \,.
 \ee
Note this effective mass-squared is a function only of the same
parameter $c$ that appears in the exponential potential for $\sigma$
and not any other combination of the individual values $c_1$ and
$c_2$.

Relative to the Hubble scale we have
\be
 \label{defetachi}
 \eta_\chi \equiv \frac{m_\chi^2}{3H^2} = - \frac23\epsilon^2
 +2\epsilon \,,
\ee
where we define the dimensionless parameter
\be
 \label{defepsilon}
\epsilon \equiv - \frac{\dot{H}}{H^2} \,,
\ee
and we have $\epsilon=1/p$ for a power-law solution, and
$\epsilon\gg1$ during ekpyrotic collapse.

Thus there is a strong tachyonic instability associated with the
ekpyrotic scaling solution (\ref{backgroundij}).

There is a simple explanation for why the scaling solution is
unstable. This represents evolution along an effective exponential
potential (\ref{Vsigmachi}) which is less steep than any of the
individual potentials (\ref{Vi}) from which it was constructed:
\be
 c^2 < c_i^2 \ \forall \ i \,.
\ee
This is what makes assisted inflation solutions stable in an expanding
universe as the field direction with the flatest potential tends to
win out at late times. In a collapsing universe, by contrast, the
field direction with the steepest (negative) gradient and hence the
stiffest equation of state tends comes to dominate at late times.

% DW
A full phase space analysis is presented in an Appendix.
In addition to the scaling solution we have fixed points
corresponding to any one of the original fields $\phi_i$ dominating
the energy density where the other fields have negligible energy
density. These correspond to power-law solutions where
\be
 p = \frac{2}{c_i^2} \,,
\ee
for $c_i^2>6$. We find that any of these single dominant field
solutions is a stable local attractor at late times during
collapse. We present the phase-space analysis in Appendix A.
%To be specific we assume that $\phi_1$ has the steepest
%potential, so that $c_1^2>c_2^2>\dots$.

Finally we note that the scaling solution (\ref{backgroundij}) is
also not the early time attractor. Restricting our attention to
solutions along the adiabatic field direction, where the entropy
fields remain in the minimum value of $U(\chi_i)$, we can use the
phase-plane analysis for a single scalar field with a steep negative
potential \cite{Heard:2002dr}. These tend to be dominated at early
times in a collapsing universe by the kinetic energy of the scalar
fields and the potential is negligible. This corresponds to a power
law solution this time with
 \be
 p=\frac13 \,.
 \ee
This is the original pre big bang proposal \cite{Gasperini:1992em}
for an early collapse phase dominated by the kinetic energy of
effectively massless moduli fields (see also
\cite{Lidsey:1999mc,Gasperini:2002bn}).

In summary, we confirm that there exists a scaling solution
corresponding to the combined evolution of two or more scalar fields
rolling down their exponential potentials. However this solution is a saddle
point in the classical phase-space of the system (see Fig.~1 in Appendix A),
representing either a set of measure zero in terms of initial conditions, or
a transient evolution between the early- and late-time attractors that
we have identified.

\section{Perturbations}

For a collapse model where $\ddot{a}<0$, small-scale fluctuations
that start far inside the Hubble length, $H^{-1}$, contract less
slowly than the Hubble length and may enter an overdamped
long-wavelength regime on super-Hubble scales, just as sub-Hubble
modes are streched up to super-Hubble scales during an inflationary
expansion. Thus initial vacuum fluctuations on small scales can
provide the initial state for large (super-Hubble) scale
perturbations during collapse.

We will consider the evolution of isocurvature modes in each of the
power-law background solutions in the previous section.

Linear perturbations in fields orthogonal to the background
trajectory are decoupled from first-order metric perturbations so
long as the trajectory remains straight in field-space. They obey
the wave equation for a massive field in an unperturbed FRW metric
 \be
 \label{isocurvatureeom}
 \ddot{\delta\chi} + 3H\dot{\delta\chi} + \left( \frac{k^2}{a^2} + m_\chi^2 \right)
 \delta \chi= 0 \,.
 \ee
Note that such isocurvature perturbations are automatically
gauge-invariant.
Introducing the rescaled field $v=a\delta\chi$, and writing the wave
equation in terms of conformal time $\tau=\int dt/a$, we have
 \be
 v'' + \left[ k^2 - \frac{a''}{a} + m_\chi^2 a^2 \right] v = 0 \,,
 \ee
where a prime denotes derivatives with respect to $\tau$.
For any power-law solution $a\propto |t|^p$ we have
 \be
 aH = \frac{1}{(\epsilon-1)\tau} \,,
 \ee
and
 \be
 \frac{a''}{a} = -(\epsilon-2)a^2H^2 \,.
 \ee
Thus for any isocurvature field we can write
 \be
 \label{v2prime}
 v'' + \left[ k^2 +
 \frac{\epsilon+3\eta_\chi-2}{(\epsilon-1)^2\tau^2} \right] v = 0
 \,,
 \ee
where we have used the dimensionless fast (or slow) roll parameters
$\epsilon=1/p$ and $\eta_\chi$, defined in Eq.~(\ref{defetachi}).

Using the usual Bunch-Davies vacuum state to normalise the amplitude
of fluctuations at early times, we obtain the
\be
v =
 \frac{\sqrt\pi}{2} \frac{e^{i(\nu+1/2)\frac{\pi}{2}}}{k^{1/2}}
 (-k\tau)^{1/2} H^{(1)}_{\nu}(-k\tau) \,, \label{vacsoln}
 \ee
where the order of the Hankel function is given by
\be
 \label{defnu}
 \nu^2 = \frac94 -
 \frac{2\epsilon^2-3\epsilon+3\eta_\chi}{(\epsilon-1)^2} \,,
 \ee
and we take $\nu\geq0$ without loss of generality.
At late times, $-k\tau\gg\nu$, this yields \cite{Allen:2004vz}
\be
 \label{Pdeltachi}
 {\cal P}_{\delta\chi} \equiv \frac{k^3}{2\pi^2} |\delta\chi^2|
 = C_\nu^2 \frac{k^2}{a^2} (-k\tau)^{1-2\nu} \,,
\ee
where $C_\nu=2^{\nu-3/2}\Gamma(\nu)/\pi^{3/2}$,
and hence we obtain a spectrum of isocurvature perturbations on large,
super-Hubble, scales with spectral tilt
\be
 \label{Deltan}
 \Delta n_{\delta\chi} \equiv \frac{d\ln {\cal P}_{\delta\chi}}{d\ln k}
 = 3-2\nu \,.
\ee

\subsection{Scaling solution}

We have seen that the isocurvature field has a tachyonic instability
with a negative effective mass-squared, given in
Eq.~(\ref{defetachi}). Substituting this into the wave equation
(\ref{v2prime}) we obtain
\be
 \nu^2 = \frac94 - \frac{3\epsilon}{(\epsilon-1)^2} \,.
\ee
Thus from Eq.~(\ref{Deltan}) we obtain a scale-invariant spectrum for
the isocurvature perturbations either as $\epsilon\to0$, corresponding
to slow-roll inflation, or as $\epsilon\to\infty$, corresponding to
ekpyrotic collapse. To leading order in a fast-roll expansion
($\epsilon\gg1$) we obtain
\be
 \Delta n_{\delta\chi} \simeq \frac{2}{\epsilon} \,.
\ee

Thus we obtain a slightly blue spectrum for a steep exponential
potential \cite{Lehners:2007ac,Buchbinder:2007ad,Creminelli:2007aq}, 
but becoming scale-invariant as
$\epsilon\to\infty$.  Any deviations from an exponential potential for
the adiabatic field, and hence the exact power-law collapse,
introduces corrections into the tilt \cite{Lehners:2007ac,Buchbinder:2007ad}, 
but we note that such corrections could also alter the effective mass
$\eta_\chi$ and the tilt becomes model-dependent.

\subsection{Single-field dominated solution}

As remarked earlier, the scaling solution during collapse
(\ref{scalingp}) is unstable with respect to collapse dominated by any
single field, $\phi_i$. In this case fluctuations in any orthogonal
field, $\phi_j$, correspond to isocurvature perturbations decoupled
from $\phi_i$ and first-order metric perturbations, so we can again
use Eq.~(\ref{isocurvatureeom}) where we identify
$\delta\chi=\delta\phi_j$. However in this case the effective mass of
the isocurvature field,
 \be
 m_\chi^2 = - c_j^2 V_j e^{-c_j\phi_j} \,,
 \ee
is much less than the Hubble rate, $H^2$. Substituting the dimensionless
mass parameter, $\eta_\chi\simeq0$, into Eq.~(\ref{v2prime}), we find
\be
 \nu^2 \simeq \frac14
% - \frac{1}{\epsilon}
 \,,
\ee
and thus the spectral tilt
\be
 \Delta n_2 = 2 \,.
\ee
We recover the same steep blue spectrum for isocurvature
perturbations during (stable) ekpyrotic collapse dominated by a
single-field, as is found for perturbations in the dominant, adiabatic
field \cite{Lyth:2001pf}.

\subsection{Kinetic-dominated solution}

We have seen that an early time attractor is the kinetic dominated
evolution with $p=1/3$ proposed in the pre-big bang scenario. In this
limit all the scalar field potentials are negligible and all field
perturbations, adiabatic and isocurvature, acquire steep blue spectra
with $\Delta n=3$, if they have canonical kinetic terms
\cite{Lidsey:1999mc}.

On the other hand there are pseudo-scalar axion fields with
non-minimal kinetic terms. Such isocurvature fields with a kinetic
coupling to dilaton-type fields can acquire a scale-invariant spectrum
of perturbations during kinetic dominated collapse as first noted in
Ref.~\cite{Copeland:1997ug} (see also \cite{Lidsey:1999mc}).

%\section{Interpretting the instability}
\section{Discussion}

We have seen that in this model we are only able to generate a
scale-invariant spectrum of isocurvature field perturbations during
collapse in the scaling solution (\ref{scalingp}) which is a saddle
point in the phase space of the system \cite{Guo:2003rs}. Thus it is
neither the early, nor late time attractor of the system.

As a result it does not solve the classical homogeneity problem.
Large scale inhomogeneneities in the initial spatial distribution of
the fields will grow during the collapse phase, in contrast to the
original single-field ekpyrotic collapse.

Even if some preceding phase drives the classical background solution
to this unstable fixed point throughout space, one might worry that
quantum fluctuations could destabilise this classical evolution.
%
%In this section we will estimate the growth of the tachyonic
%instability.
%
We see from Eq.~(\ref{Pdeltachi}) that for long wavelength
perturbations we have $\delta\chi\propto
\tau^{1-2\nu}/a^2$ and in the ekpyrotic scaling solution where
$\nu\sim3/2$ this gives $\delta\chi \sim H$.

Indeed it is easy to see that we always require an instability of
this form
% DW
in a canonical scalar field in order to generate a scale invariant
spectrum during collapse, as the amplitude of field perturbations at
Hubble exit are also of order $H$ and thus we require the
super-Hubble perturbations to grow at the same rate to maintain a
scale-invariant spectrum. Precisely the same form of instability
appears in the power-law collapse model with $p=2/3$
% DW
which produces a scale-invariant spectrum of comoving curvature
perturbations from quantum fluctuations
\cite{Wands:1998yp,Finelli:2001sr,Gratton:2003pe,Allen:2004vz}.

As $H$ is rapidly growing in the ekpyrotic scenario ($\epsilon\gg1$)
this corresponds to a rapid tachyonic growth of the entropy field.
On the other hand one should remember that the minimum duration of
such a phase is also set by the rate of change of $H$. The change in
the Hubble rate $H$ determiness the range of comoving scales that
exit the Hubble scale (as the scale factor $a$ changes only slowly).
Thus it might be sufficient for $|H|$ and thus $\delta\chi$ to grow
by a factor of $k_{\rm max}/k_{\rm min}\sim 10^5$ while observable
scales exit the Hubble scale.

The actual density of long wavelength quantum fluctuations (per
logarithmic range in wavelength) either from the kinetic energy or
potential energy density of the entropy field fluctuations, can be
estimated relative to the total energy density to be
\be
\Omega_\chi \sim \epsilon H^2 \,.
\ee
Thus the fractional density of the entropy modes remains small and
does not threaten to destabilise the classical evolution so long as
the Hubble rate remains well below the Planck scale (here set to
unity).

In any case we do not necessarily expect the scaling solution, or any
classical solution to hold arbitrarily close to the Planck scale,
especially if we want to appeal to high-energy corrections to produce
a bounce in the scale-factor, as envisaged in pre big bang models
\cite{Gasperini:1992em,Gasperini:2002bn} and now proposed in ekpyrotic models
\cite{Buchbinder:2007ad,Creminelli:2007aq}.

We note that the existence of a natural turning point in the
field-space trajectory due to the instability of the scaling
solution offers the possibility of converting the scale-invariant
spectrum of isocurvature field perturbations into a scale-invariant
spectrum of curvature perturbations before the singularity or bounce
without introducing any additional mechansim
\cite{Lehners:2007ac,Buchbinder:2007ad,Creminelli:2007aq}.

Thus the ekpyrotic collapse with multiple fields offers an alternative
scenario for the origin of structure in our Universe to contrast with
inflationary models. However to do so we seem to require at least four
distinct phases in the evolution: (1) an initial phase to establish
spatial homogeneity, (2) a scaling collapse phase to produce a
scale-invariant spectrum of field perturbations, (3) a phase to turn
isocurvature field perturbations into curvature perturbations, and (4)
a bounce phase to turn collapse into expansion. Note that phases (3)
and (4) could happen in the reverse order, as in the curvaton
mechanism. By comparison, the simplest inflationary models require
only a single quasi-de Sitter phase to establish spatial homogeneity
plus a scale-invariant spectrum of curvature perturbations.

%\section{Conclusions}

\appendix
\section{Appendix}

In this appendix, we present the phase space analysis for
mutli-fields. Note that a similar analysis was done in
Ref.~\cite{Guo:2003rs} for the case $c_1=c_2$ in an expanding
universe. The field equations and an acceleration equation are
\begin{equation}
\phi_i'' + 2 h \phi_i' +a^2 c_i V_i e^{-c_i \phi_i} =0,
\end{equation}
\begin{equation}
h' = \frac{1}{3}\sum_i(- \phi_i^{'2} - a^2 V_i e^{-c_i \phi_i}),
\end{equation}
where prime denotes a derivative with respect to a conformal
time and $h=a'/a$. The Friedman equation is given by
\begin{equation}
h^2= \frac{1}{3} \sum_i \left(\frac{1}{2} \phi_i^{'2}
- a^2 V_i e^{-c_i \phi_i}\right).
\end{equation}
We define the dimensionless phase-space variables
\begin{eqnarray}
x_i &=& \frac{\phi_i'}{\sqrt{6} h}, \\
y_i &=& \frac{a \sqrt{V_i e^{-c_i \phi_i}}}{\sqrt{3} h}.
\end{eqnarray}
As we are interested in a contracting universe, $y_i <0$.
The field equations and the evolution equation for $a$ give the
first order evolution equations for the phase space variables
\begin{eqnarray}
\frac{d x_i}{d N} &=& -3 x_i (1- \sum_j x_j^2) - c_i \sqrt{\frac{3}{2}} y_i^2, \\
\frac{d y_i}{d N} &=& y_i \left(3 \sum x_j^2  - c_i \sqrt{\frac{3}{2}} x_i \right),\\
\end{eqnarray}
where $N = \log a$.
The Friedmann equation gives a constraint
\begin{equation}
\sum_j x_j^2 - \sum_j y_j^2 =1.
\end{equation}

There are $n+2$ fixed points of the system where $dx_i/dN = dy_i/dN
=0$.
\begin{eqnarray}
A : &&\;\;\; \sum_j x_j^2 =1, \quad y_j =0.\\
B_i : &&\;\;\; x_i =\frac{c_i}{\sqrt{6}}, \;\;\;
y_i=-\sqrt{\frac{c_i^2}{6}-1},
 \;\;\;
x_j=y_j=0, \;\; (\mbox{for} \;\;\; j \neq i),\\
B : &&\;\;\; x_j = \frac{\sqrt{6}}{3 p} \frac{1}{c_j}, \;\;\; y_j =
\sqrt{\frac{2}{c_j^2 p} \left(\frac{1}{3p}-1 \right) },
\end{eqnarray}
where
\begin{equation}
p = \sum_j \frac{2}{c^2_j}.
\end{equation}

$A$ is a series of kinetic dominated solutions where the potential
energy can be neglected. $B_i$ is a single field scaling solution
where one of the fields is dominating the energy density. $B$ is the
scaling solution which is realized by the combined evolution of
multiple fields. Note that at the fixed point $B$, the solution for
the scale factor is given by
\begin{equation}
a = (-t)^p = (- \tau)^{-p/(p-1)}.
\end{equation}

To study the stability of the solutions, the linear perturbations
around each fixed point have to be analyzed. For simplicity we
restrict ourselves to the case of two fields. We can eliminate $y_2$
using the constraint $y_2^2 = -1 +x_1^2+x_2^2 -y_1^2$. Then the
phase space becomes three-dimensional space spanned by
$(x_1,x_2,y_1)$. There are four fixed points $A, B_1, B_2$ and $B$.
The solution for the linearized perturbations are characterized by
three eigenfunctions:
\begin{equation}
\delta x = u_1 \exp(m_1 N) + u_2 \exp (m_2 N) + u_3 \exp(m_3 N).
\end{equation}
Stability during collapse ($N\to-\infty$) requires that the real
parts of all the eigenvalues $m_i$ are positive. The eigenvalues are
given by
\begin{eqnarray}
A: \;\;\; m_1 &=& 0, \;\; m_2 =3 -\frac{\sqrt{6}}{2} c_1 x_1,
\;\; m_3 =6 -\sqrt{6} c_2 x_2,\\
B_1: \;\;\; m_1 &=& c_1^2, \;\; m_2 = -3 + \frac{c_1^2}{2},
\;\; m_3 = -3 + \frac{c_1^2}{2},\\
B_2: \;\;\; m_1 &=& \frac{c_2^2}{2}, \;\; m_2 = -3 + \frac{c_2^2}{2}, \;\;
m_3 = -3 + \frac{c_2^2}{2},\\
B : \;\;\; m_1 &=& -\frac{1}{p}(3p-1), \;\; m_2 =m_+, \;\;
m_3 = m_-,
\end{eqnarray}
where
\begin{equation}
m_{\pm} = \frac{1}{2p}(3p-1)
\left(-1 \mp \sqrt{\frac{3(p-3)}{3p-1}} \right).
\end{equation}

The kinetic-dominated solutions $A$ are unstable for $c_1 x_1 >
\sqrt{6}$ and $c_2 x_2 > \sqrt{6}$. The single field scaling
solutions, $B_i$, exist for $c_i > \sqrt{6}$ and it is always
stable. The scaling solution $B$ exists for $p<1/3$ and there is
always one negative eigenvalue, which means that the scaling
solution is a saddle point.
The instability mode is given by
\begin{equation}
\delta x \propto  e^{m_{-} N} \propto (-t)^{p m_-}.
\end{equation}
For small $p \ll 1$, $m_- = - 1/p$, thus $\delta x \propto
(-t)^{-1} \propto H$. This is exactly the instability that the
entropy field exhibits around the scaling solution. In fact, in terms of $\epsilon =1/p$,
$m_-$ is written as
\begin{equation}
m_- = -\frac{1}{2}(3 -\epsilon) +
\frac{1}{2} \sqrt{3(3 \epsilon-1)(\epsilon-3)}.
\end{equation}
Then we can show that the solutions for the entropy field Eq.~(\ref{vacsoln}) is
given by
\begin{equation}
\delta \chi \propto e^{N m_{-}}.
\end{equation}

Figure~1 shows the phase space trajectories around $B$. We see that
all solutions go to the stable fixed points $B_1$ or $B_2$ at late
times. It is clear that $B$ is a saddle point and any solutions near
$B$ go to $B_1$ or $B_2$ at late times. This is the origin of the
instability of the scaling solution.

\begin{figure}[http]
 \begin{center}
\includegraphics[width=10cm]{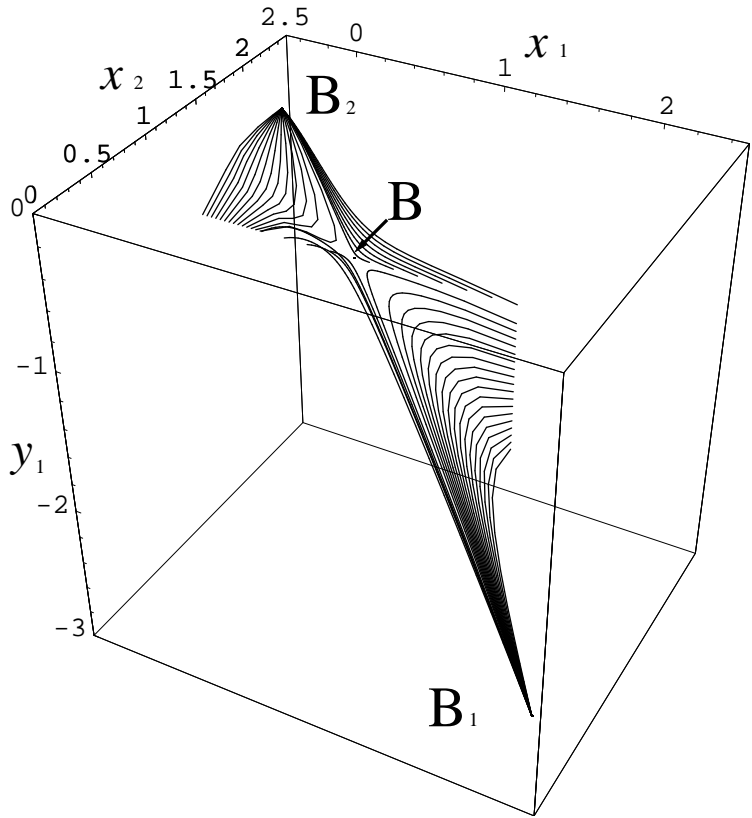}
\caption[]{Phase space trajectories around the fixed point B in
three dimensional phase space spanned by $(x_1,x_2, y_1)$. We see
that the single field fixed points $B_1$ and $B_2$ are late time
attractors and the scaling solution $B$ is a saddle point. We take
$c_1 =6$ and $c_2 =4$.Then the fixed points are given by $B_1:
(2.45, 0, -2.24)$, $B_2: (0, 1.63, 0)$ and $B: (0.75, 1.13,
-0.51)$.}
\end{center}
\end{figure}

% References


\begin{thebibliography}{99}

%\cite{Gasperini:1992em}
\bibitem{Gasperini:1992em}
  M.~Gasperini and G.~Veneziano,
  %``Pre - big bang in string cosmology,''
  Astropart.\ Phys.\  {\bf 1}, 317 (1993)
  [arXiv:hep-th/9211021].
  %%CITATION = APHYE,1,317;%%


  %\cite{Gasperini:1993hu}
\bibitem{Gasperini:1993hu}
  M.~Gasperini and G.~Veneziano,
  %``Inflation, deflation, and frame independence in string cosmology,''
  Mod.\ Phys.\ Lett.\  A {\bf 8}, 3701 (1993)
  [arXiv:hep-th/9309023].
  %%CITATION = MPLAE,A8,3701;%%


\bibitem{Kunze:1999xp}
  K.~E.~Kunze and R.~Durrer,
  %``Anisotropic 'hairs' in string cosmology,''
  Class.\ Quant.\ Grav.\  {\bf 17}, 2597 (2000)
  [arXiv:gr-qc/9912081].
  %%CITATION = CQGRD,17,2597;%%

\bibitem{Khoury:2001wf}
  J.~Khoury, B.~A.~Ovrut, P.~J.~Steinhardt and N.~Turok,
  %``The ekpyrotic universe: Colliding branes and the origin of the hot big
  %bang,''
  Phys.\ Rev.\  D {\bf 64}, 123522 (2001)
  [arXiv:hep-th/0103239].
  %%CITATION = PHRVA,D64,123522;%%

\bibitem{Kallosh:2001ai}
  R.~Kallosh, L.~Kofman and A.~D.~Linde,
  %``Pyrotechnic universe,''
  Phys.\ Rev.\  D {\bf 64}, 123523 (2001)
  [arXiv:hep-th/0104073].
  %%CITATION = PHRVA,D64,123523;%%

\bibitem{Khoury:2001iy}
  J.~Khoury, B.~A.~Ovrut, P.~J.~Steinhardt and N.~Turok,
  %``A brief comment on 'The pyrotechnic universe',''
  arXiv:hep-th/0105212.
  %%CITATION = HEP-TH/0105212;%%

 \bibitem{Lyth:2001pf}
  D.~H.~Lyth,
  %``The primordial curvature perturbation in the ekpyrotic universe,''
  Phys.\ Lett.\  B {\bf 524}, 1 (2002)
  [arXiv:hep-ph/0106153].
  %%CITATION = PHLTA,B524,1;%%

\bibitem{Brandenberger:2001bs}
  R.~Brandenberger and F.~Finelli,
  %``On the spectrum of fluctuations in an effective field theory of the
  %ekpyrotic universe,''
  JHEP {\bf 0111}, 056 (2001)
  [arXiv:hep-th/0109004].
  %%CITATION = JHEPA,0111,056;%%

\bibitem{Hwang:2001ga}
  J.~c.~Hwang,
  %``Cosmological structure problem in the ekpyrotic scenario,''
  Phys.\ Rev.\  D {\bf 65}, 063514 (2002)
  [arXiv:astro-ph/0109045].
  %%CITATION = PHRVA,D65,063514;%%



\bibitem{Khoury:2001zk}
  J.~Khoury, B.~A.~Ovrut, P.~J.~Steinhardt and N.~Turok,
  %``Density perturbations in the ekpyrotic scenario,''
  Phys.\ Rev.\  D {\bf 66}, 046005 (2002)
  [arXiv:hep-th/0109050].
  %%CITATION = PHRVA,D66,046005;%%




\bibitem{Wands:1998yp}
  D.~Wands,
  %``Duality invariance of cosmological perturbation spectra,''
  Phys.\ Rev.\  D {\bf 60}, 023507 (1999)
  [arXiv:gr-qc/9809062].
  %%CITATION = PHRVA,D60,023507;%%

  \bibitem{Finelli:2001sr}
  F.~Finelli and R.~Brandenberger,
  %``On the generation of a scale-invariant spectrum of adiabatic  fluctuations
  %in cosmological models with a contracting phase,''
  Phys.\ Rev.\  D {\bf 65}, 103522 (2002)
  [arXiv:hep-th/0112249].
  %%CITATION = PHRVA,D65,103522;%%
\bibitem{Allen:2004vz}
  L.~E.~Allen and D.~Wands,
  %``Cosmological perturbations through a simple bounce,''
  Phys.\ Rev.\  D {\bf 70}, 063515 (2004)
  [arXiv:astro-ph/0404441].
  %%CITATION = PHRVA,D70,063515;%%
\bibitem{Heard:2002dr}
  I.~P.~C.~Heard and D.~Wands,
  %``Cosmology with positive and negative exponential potentials,''
  Class.\ Quant.\ Grav.\  {\bf 19}, 5435 (2002)
  [arXiv:gr-qc/0206085].
  %%CITATION = CQGRD,19,5435;%%

\bibitem{Gratton:2003pe}
  S.~Gratton, J.~Khoury, P.~J.~Steinhardt and N.~Turok,
  %``Conditions for generating scale-invariant density perturbations,''
  Phys.\ Rev.\  D {\bf 69}, 103505 (2004)
  [arXiv:astro-ph/0301395].
  %%CITATION = PHRVA,D69,103505;%%


\bibitem{Copeland:1997ug}
  E.~J.~Copeland, R.~Easther and D.~Wands,
  %``Vacuum fluctuations in axion-dilaton cosmologies,''
  Phys.\ Rev.\  D {\bf 56}, 874 (1997)
  [arXiv:hep-th/9701082].
  %%CITATION = PHRVA,D56,874;%%

\bibitem{Lidsey:1999mc}
  J.~E.~Lidsey, D.~Wands and E.~J.~Copeland,
  %``Superstring cosmology,''
  Phys.\ Rept.\  {\bf 337}, 343 (2000)
  [arXiv:hep-th/9909061].
  %%CITATION = PRPLC,337,343;%%

\bibitem{Enqvist:2001zp}
  K.~Enqvist and M.~S.~Sloth,
  %``Adiabatic CMB perturbations in pre big bang string cosmology,''
  Nucl.\ Phys.\  B {\bf 626}, 395 (2002)
  [arXiv:hep-ph/0109214].
  %%CITATION = NUPHA,B626,395;%%

\bibitem{Lyth:2001nq}
  D.~H.~Lyth and D.~Wands,
  %``Generating the curvature perturbation without an inflaton,''
  Phys.\ Lett.\  B {\bf 524}, 5 (2002)
  [arXiv:hep-ph/0110002].
  %%CITATION = PHLTA,B524,5;%%



\bibitem{Notari:2002yc}
  A.~Notari and A.~Riotto,
  %``Isocurvature perturbations in the ekpyrotic universe,''
  Nucl.\ Phys.\  B {\bf 644}, 371 (2002)
  [arXiv:hep-th/0205019].
  %%CITATION = NUPHA,B644,371;%%

\bibitem{DiMarco:2002eb}
  F.~Di Marco, F.~Finelli and R.~Brandenberger,
  %``Adiabatic and Isocurvature Perturbations for Multifield Generalized
  %Einstein Models,''
  Phys.\ Rev.\  D {\bf 67}, 063512 (2003)
  [arXiv:astro-ph/0211276].
  %%CITATION = PHRVA,D67,063512;%%

\bibitem{Finelli:2002we}
  F.~Finelli,
  %``Assisted contraction,''
  Phys.\ Lett.\  B {\bf 545}, 1 (2002)
  [arXiv:hep-th/0206112].
  %%CITATION = PHLTA,B545,1;%%

\bibitem{Guo:2003rs}
  Z.~K.~Guo, Y.~S.~Piao, R.~G.~Cai and Y.~Z.~Zhang,
  %``Cosmological scaling solutions and cross-coupling exponential  potential,''
  Phys.\ Lett.\  B {\bf 576}, 12 (2003)
  [arXiv:hep-th/0306245].
  %%CITATION = PHLTA,B576,12;%%

\bibitem{Liddle:1998jc}
  A.~R.~Liddle, A.~Mazumdar and F.~E.~Schunck,
  %``Assisted inflation,''
  Phys.\ Rev.\  D {\bf 58}, 061301 (1998)
  [arXiv:astro-ph/9804177].
  %%CITATION = PHRVA,D58,061301;%%


\bibitem{Lehners:2007ac}
  J.~L.~Lehners, P.~McFadden, N.~Turok and P.~J.~Steinhardt,
  %``Generating ekpyrotic curvature perturbations before the big bang,''
  arXiv:hep-th/0702153.
  %%CITATION = HEP-TH/0702153;%%

\bibitem{Buchbinder:2007ad}
  E.~I.~Buchbinder, J.~Khoury and B.~A.~Ovrut,
  %``New ekpyrotic cosmology,''
  arXiv:hep-th/0702154.
  %%CITATION = HEP-TH/0702154;%%

\bibitem{Creminelli:2007aq}
  P.~Creminelli and L.~Senatore,
  %``A smooth bouncing cosmology with scale invariant spectrum,''
  arXiv:hep-th/0702165.
  %%CITATION = HEP-TH/0702165;%%



\bibitem{Gasperini:1996fu}
  M.~Gasperini, M.~Maggiore and G.~Veneziano,
  %``Towards a non-singular pre-big bang cosmology,''
  Nucl.\ Phys.\  B {\bf 494}, 315 (1997)
  [arXiv:hep-th/9611039].
  %%CITATION = NUPHA,B494,315;%%

\bibitem{Brustein:1997ny}
  R.~Brustein and R.~Madden,
  %``Graceful exit and energy conditions in string cosmology,''
  Phys.\ Lett.\  B {\bf 410}, 110 (1997)
  [arXiv:hep-th/9702043].
  %%CITATION = PHLTA,B410,110;%%

\bibitem{Brustein:1997cv}
  R.~Brustein and R.~Madden,
  %``A model of graceful exit in string cosmology,''
  Phys.\ Rev.\  D {\bf 57}, 712 (1998)
  [arXiv:hep-th/9708046].
  %%CITATION = PHRVA,D57,712;%%

\bibitem{Creminelli:2006xe}
  P.~Creminelli, M.~A.~Luty, A.~Nicolis and L.~Senatore,
  %``Starting the universe: Stable violation of the null energy condition and
  %non-standard cosmologies,''
  JHEP {\bf 0612}, 080 (2006)
  [arXiv:hep-th/0606090].
  %%CITATION = JHEPA,0612,080;%%


\bibitem{Tsujikawa:2002qc}
  S.~Tsujikawa, R.~Brandenberger and F.~Finelli,
  %``On the construction of nonsingular pre-big-bang and ekpyrotic cosmologies
  %and the resulting density perturbations,''
  Phys.\ Rev.\  D {\bf 66}, 083513 (2002)
  [arXiv:hep-th/0207228].
  %%CITATION = PHRVA,D66,083513;%%


\bibitem{Gordon:2000hv}
  C.~Gordon, D.~Wands, B.~A.~Bassett and R.~Maartens,
  %``Adiabatic and entropy perturbations from inflation,''
  Phys.\ Rev.\  D {\bf 63}, 023506 (2001)
  [arXiv:astro-ph/0009131].
  %%CITATION = PHRVA,D63,023506;%%

\bibitem{Malik:1998gy}
  K.~A.~Malik and D.~Wands,
  %``Dynamics of assisted inflation,''
  Phys.\ Rev.\  D {\bf 59}, 123501 (1999)
  [arXiv:astro-ph/9812204].
  %%CITATION = PHRVA,D59,123501;%%



\bibitem{Gasperini:2002bn}
  M.~Gasperini and G.~Veneziano,
  %``The pre-big bang scenario in string cosmology,''
  Phys.\ Rept.\  {\bf 373}, 1 (2003)
  [arXiv:hep-th/0207130].
  %%CITATION = PRPLC,373,1;%%



\end{thebibliography}
\end{document}